\documentclass[aps,prb,twocolumn, amsmath, amssymb,showpacs,superscriptaddress]{revtex4-1}

\usepackage{graphicx}
\usepackage{amsmath,amssymb,amsfonts}
\usepackage{textcomp}
\usepackage{hyperref}
\usepackage{gensymb}
\usepackage{verbatim}
\usepackage{amsmath}
\hypersetup{breaklinks=true,colorlinks=true,urlcolor=black}
\usepackage{color}
\usepackage{soul}
\DeclareGraphicsExtensions{.jpg,.pdf,.png,.eps}

\begin{document}
\title{Extremely large magnetoresistance and Kohler's rule in PdSn$_4$: a complete study of thermodynamic, transport and band structure properties.}
\author{Na Hyun Jo}
\affiliation{Ames Laboratory, Iowa State University, Ames, Iowa 50011, USA}
\affiliation{Department of Physics and Astronomy, Iowa State University, Ames, Iowa 50011, USA}

\author{Yun Wu}
\affiliation{Ames Laboratory, Iowa State University, Ames, Iowa 50011, USA}
\affiliation{Department of Physics and Astronomy, Iowa State University, Ames, Iowa 50011, USA}

\author{Lin-Lin Wang}
\affiliation{Ames Laboratory, Iowa State University, Ames, Iowa 50011, USA}

\author{Peter P. Orth}
\affiliation{Ames Laboratory, Iowa State University, Ames, Iowa 50011, USA}
\affiliation{Department of Physics and Astronomy, Iowa State University, Ames, Iowa 50011, USA}

\author{Savannah S. Downing}
\affiliation{Ames Laboratory, Iowa State University, Ames, Iowa 50011, USA}
\affiliation{Department of Physics and Astronomy, Iowa State University, Ames, Iowa 50011, USA}

\author{Soham Manni}
\affiliation{Ames Laboratory, Iowa State University, Ames, Iowa 50011, USA}
\affiliation{Department of Physics and Astronomy, Iowa State University, Ames, Iowa 50011, USA}

\author{Dixiang Mou}
\affiliation{Ames Laboratory, Iowa State University, Ames, Iowa 50011, USA}
\affiliation{Department of Physics and Astronomy, Iowa State University, Ames, Iowa 50011, USA}

\author{Duane D. Johnson}
\affiliation{Ames Laboratory, Iowa State University, Ames, Iowa 50011, USA}
\affiliation{Department of Physics and Astronomy, Iowa State University, Ames, Iowa 50011, USA}
\affiliation{Department of Materials Science and Engineering, Iowa State University, Ames, Iowa 50011, USA}

\author{Adam Kaminski}
\affiliation{Ames Laboratory, Iowa State University, Ames, Iowa 50011, USA}
\affiliation{Department of Physics and Astronomy, Iowa State University, Ames, Iowa 50011, USA}

\author{Sergey L. Bud'ko}
\affiliation{Ames Laboratory, Iowa State University, Ames, Iowa 50011, USA}
\affiliation{Department of Physics and Astronomy, Iowa State University, Ames, Iowa 50011, USA}

\author{Paul C. Canfield}
\affiliation{Ames Laboratory, Iowa State University, Ames, Iowa 50011, USA}
\affiliation{Department of Physics and Astronomy, Iowa State University, Ames, Iowa 50011, USA}
\email[]{canfield@ameslab.gov}

\date{\today}

\begin{abstract}
The recently discovered material PtSn$_4$ is known to exhibit extremely large magnetoresistance (XMR) that also manifests Dirac arc nodes on the surface. PdSn$_4$ is isostructure to PtSn$_4$ with same electron count. We report on the physical properties of high quality single crystals of PdSn$_4$ including specific heat, temperature and magnetic field dependent resistivity and magnetization, and electronic band structure properties obtained from angle resolved photoemission spectroscopy (ARPES). We observe that PdSn$_4$ has physical properties that are qualitatively similar to those of PtSn$_4$, but find also pronounced differences. Importantly, the Dirac arc node surface state of PtSn$_4$ is gapped out for PdSn$_4$. By comparing these similar compounds, we address the origin of the extremely large magnetoresistance in PdSn$_4$ and PtSn$_4$; based on detailed analysis of the magnetoresistivity, $\rho(H,T)$, we conclude that neither carrier compensation nor the Dirac arc node surface state are primary reason for the extremely large magnetoresistance. On the other hand, we find that surprisingly Kohler's rule scaling of the magnetoresistance, which describes a self-similarity of the field induced orbital electronic motion across different length scales and is derived for a simple electronic response of metals to applied in a magnetic field is obeyed over the full range of temperatures and field strengths that we explore.
\end{abstract}
\date{\today}
\maketitle 

\section{Introduction}
The recent discovery of three-dimensional topological semi-metals generated lots of attention. Topological semi-metals can be classified as three different types:\cite{Armitage2017} a Dirac semimetal, a Weyl semimetal, and a nodal-line semimetal. A Dirac semimetal has points in the Brillouin zone where doubly degenerate conduction and valence bands are touching linearly, and has been realized in Na$_3$Bi and Cd$_3$As$_2$.\cite{Liu2014,Wang2013} A Weyl semimetal is similar to Dirac semimetal, but a Weyl semi-metal exhibits Weyl points that are crossing of singly degenerate electronic bands, which requires the breaking of either time-reversal or inversion symmetry. Weyl points have chirality which is represented by the sign of the determinant for the velocity tensor.\cite{Weng2016} Experimental manifestations of Weyl semimetals are: TaAs, NbAs, TaP and transition metal dichalcogenides including WTe$_2$ .\cite{Lv2015,Xu2015,XuAli2015,XuBel2015,WuMouJo2016} A nodal-line semimetal has a gap closing along a line in the Brillouin zone instead of at isolated points. Several materials have experimentally confirmed to be nodal-line semimetals: PbTaSe$_2$, ZrSiS and TlTaSe$_2$.\cite{Bian2016,Schoop2016,BianChang2016} A different type of a novel topological quantum structure was identified recently in PtSn$_4$ which has a Dirac node arc feature on the surface, whose (topological) origin is still unknown.\cite{WuPtSn2016} In the bulk, PtSn$_4$ is a metallic with a number of complex electron and hole Fermi surface sheets containing significant carrier densities.\cite{WuPtSn2016} 

The fascinating physical properties of PtSn$_4$ were reported in Ref.\,\onlinecite{Mun2012}. Most importantly, it manifests an extremely large magnetoresistance (XMR) of $\sim$\,10$^{5}\,\%$ for $T\,=\,1.8$\,K and $H\,=\,140$\,kOe, without any evidence of saturation.\cite{Mun2012} In the following years, many of other topological materials that also showed XMR have been found. 

Several mechanisms have been suggested for such XMR in various materials. Nearly perfect electron-hole compensation has been suggested as the origin of XMR ($\sim\,10^{5}\,\%$ at 4.5\,K and 147\,kOe) in WTe$_2$.\cite{Ali2014} (Although it has also been noted that in this material there is a clear relation between MR and the residual resistivity ratio, $\rho_{0}$)\cite{ali2015}. Magnetic field induced changes in Fermi surface structure have been suggested as the cause of XMR in NbSb$_2$ ($\sim\,10^{5}\,\%$ at 2\,K and 9\,T) and Cd$_3$As$_2$ ($\sim\,10^{5}\,\%$ at 5\,K and 9\,T), specifically field induced gaps in Dirac points or splitting of Dirac Fermi points into separated Weyl points.\cite{Ali2014,Liang2015} More recently, for the case of LaSb ($\sim\,10^{6}\,\%$ at 2\,K and 9\,T) and LaBi ($\sim\,10^{5}\,\%$ at 2\,K and 9\,T) XMR has been associated with $d$-$p$ orbital mixing.\cite{TaftiGibson2016,tafti2016} Despite all of these possible mechanisms, the origins of XMR in all of these materials are open to debate. 

PdSn$_4$ is isostructural to PtSn$_4$;\cite{Kubiak1984,Nylen2004} it has an orthorhombic structure and a space group number of 68, with Pd and Sn each having a unique crystallographic site.\cite{Nylen2004} The lattice parameters of PdSn$_4$ are $a\,=\,6.4417\,\AA$, $b\,=\,11.4452\,\AA$ and $c\,=\,6.3886\,\AA$, which are very similar to those of PtSn$_4$: $a\,=\,6.418\,\AA$, $b\,=\,11.366\,\AA$ and $c\,=\,6.384\,\AA$.\cite{Nylen2004,Kuennen2000} Since not only the structure, but also the electron count is same, the primary difference between PdSn$_4$ and PtSn$_4$ may be the spin orbit coupling strength. Therefore, it is interesting to compare these two materials. 

In this paper, we report the physical properties of PdSn$_4$, including temperature and the magnetic field dependent transport and magnetic properties as well as a detailed angle resolved photoemmision spectroscopy (ARPES) and density functional theory (DFT) study of its electronic bulk and surface band structure. Compared to PtSn$_4$, the anisotropic physical properties of PdSn$_4$ are qualitatively similar. In contrast, our ARPES results show that whereas PtSn$_4$ exhibits a Dirac node arc at the surface, the corresponding surface state is clearly gapped out in PdSn$_4$. We rule out a number of proposed microscopic mechanisms for XMR, through detailed analysis of the magnetoresistivity and band structure. In particular, we show that carrier compensation cannot explain XMR in PtSn$_4$ and PdSn$_4$. In addition, we point out a surprising scaling property of the magnetoresistivity that follows Kohler's rule over a wide magnetic field and temperature range, revealing to a robust scale invariance of the electronic orbital motion known from classical descriptions of magnetotransport.    

\section{experimental and theoretical methods}

\begin{figure}
	\includegraphics[scale=1]{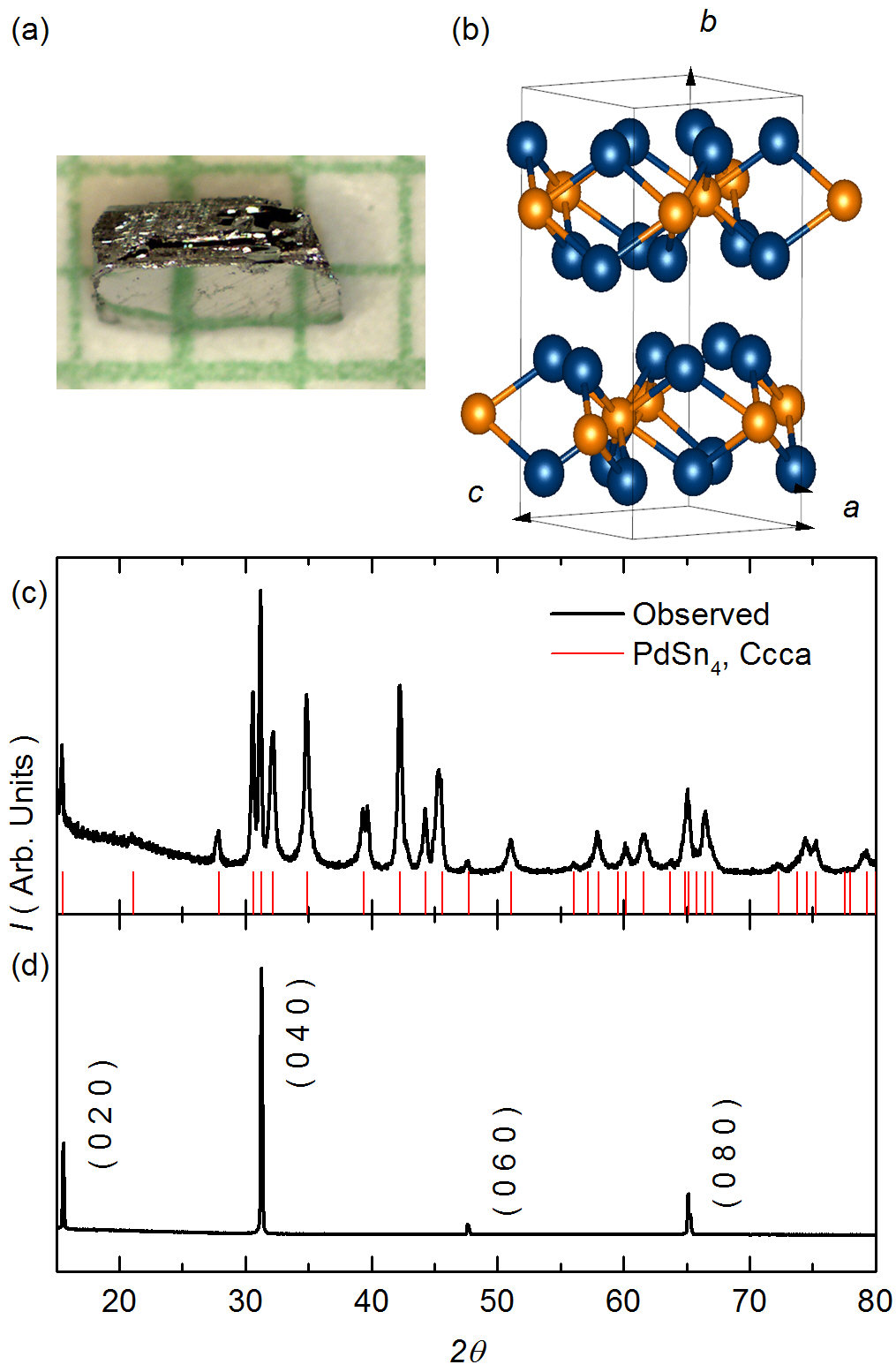}%
	\caption{(color online) (a) Single crystal of PdSn$_4$. (b) Crystal structure (Pd: orange sphere, Sn: blue sphere). (c) Powder X-ray diffraction pattern (Black lines) and reported PdSn$_4$ ($Ccca$) peaks from ref.\cite{Nylen2004} (Red lines). (d) Single crystal X-ray diffraction pattern identifying the (0 h 0) lines. 
		\label{xrd}}
\end{figure}

Single crystals of PdSn$_4$ were grown out of Sn-rich binary melts.\cite{ASM2000} We put an initial stoichiometry of Pd$_2$Sn$_{98}$ into a fritted alumina crucible [CCS],\cite{Canfield2016} and then sealed in it into an amorphous silica tube under partial Ar atmosphere. The ampoule was heated up to 600\,\celsius\ over 5\,hours, held there for 5\,hours, rapidly cooled to 325\,\celsius\, and then slowly cooled down to 245\,\celsius\ over more than 100\,hours, and then finally decanted using a centrifuge.\cite{Canfield1992} The fritted crucible allowed for the clean separation of crystals from the Sn flux.\cite{Petrovic2012} The single crystalline sample has a clear plate like shape with mirrored surface and typical dimension of $\sim$\,2\,mm\,$\times$\,1\,mm $\times$\,0.3\,mm.(Fig.\,\ref{xrd} (a))

A Rigaku MiniFlex diffractometer (Cu $K_{\alpha1,2}$ radiation) was used for acquiring x-ray diffraction (XRD) data at room temperature (Fig.\,\ref{xrd} (c)). All the peak positions are well matched with the reported orthorhombic PdSn$_4$ structure.\cite{Nylen2004} However, the relative intensity of the peaks is different from the calculated powder pattern, presumably preferential orientation of the ground powder due to the layered structure. The crystal structure in real space is shown in Fig.\,\ref{xrd} (b). The crystallographic $b$-axis is anticipated to be perpendicular to the mirrored surface shown in Fig.\,\ref{xrd} (a). To confirm this, we did single crystal XRD.\cite{Jesche2016} When the x-ray beam is incident perpendicular to the surface, only (0 k 0) peaks were observed as shown in Fig.\,\ref{xrd} (d). This confirms that the crystallographic $b$ direction is perpendicular to the mirrored surface shown in Fig.\,\ref{xrd} (a). 

Temperature and field dependent transport properties were measured in a Quantum Design(QD), Physical Property Measurement System for 1.8\,$\le T \le$\,300\,K and $\left| H \right| \le$ \,140\,kOe. The contacts for the electrical transport measurement were prepared in a standard four-probe configuration using Epotek-H20E silver epoxy. The magnetization as a function of temperature and field was measured in a QD, Magnetic Property Measurement System for 1.85\,$\le T \le$\,300\,K and $H \le$\,70\,kOe. A modified QD sample rotating platform with an angular resolution of 0.1\,$\deg$ was used for angular-dependent dc magnetization measurements. 

ARPES measurements were carried out using a laboratory-based tunable VUV laser ARPES system, consisting of a Scienta R8000 electron analyzer, picosecond Ti:Sapphire oscillator and fourth harmonic generator.\cite{Jiang2014} All Data were collected with a constant photon energy of 6.7 eV. Angular resolution was set at $\sim$ 0.05$^\circ$ and 0.5$^\circ$ (0.005~\AA$^{-1}$ and 0.05 ~\AA$^{-1}$) along and perpendicular to the direction of the analyzer slit (and thus cut in the momentum space), respectively; and energy resolution was set at 1~meV. The size of the photon beam on the sample was $\sim$30~$\mu$m. Samples were cleaved \textit{in situ} at a base pressure lower than $1 \times 10^{-10}$~Torr. Samples were cleaved at 40~K and kept at the cleaving temperature throughout the measurement. 

Band structure with spin-orbital coupling (SOC) in density functional theory (DFT)\cite{Hohenberg1964,Kohn1965} have been calculated with PBE\cite{Perdew1996} exchange-correlation functional, a plane-wave basis set and projected augmented wave method\cite{Bloechl1994} as implemented in VASP5,6.\cite{Kresse1996,Kresse1996a} For bulk band structure of PdSn$_4$, we used the conventional orthorhombic cell of 20 atoms with a Monkhorst-Pack\cite{Monkhorst1976} (8\,$\times$\,6\,$\times$\,8) $k$-point mesh including the $\Gamma$ point and a kinetic energy cutoff of 251\,eV. The convergence with respect to $k$-point mesh was carefully checked, with total energy converged below 1 meV/atom. Experimental lattice parameters have been used with atoms fixed in their bulk positions. A tight-binding model based on maximally localized Wannier functions\cite{Marzari1997,Souza2001,Marzari2012} was constructed to reproduce closely the bulk band structure including SOC in the range of $E_{F}\,\pm\,1$\,eV with Pd $s$ and $d$ orbitals and Sn $s$ and $p$ orbitals. Then Fermi surface and spectral functions of a semi-infinite PdSn$_4$ (0 1 0) surface were calculated with the surface Green’s function methods\cite{Lee1981,LeeJoannopoulos1981,Sancho1984,Sancho1985} as implemented in WannierTools.\cite{WuQS2017} 

The PtSn$_4$ data that are shown here for comparison are from our previous paper on PtSn$_4$.\cite{Mun2012}

\section{Results and analysis}

\subsection{\texorpdfstring{Specific heat}{space}}

\begin{figure}
	\includegraphics[scale=1]{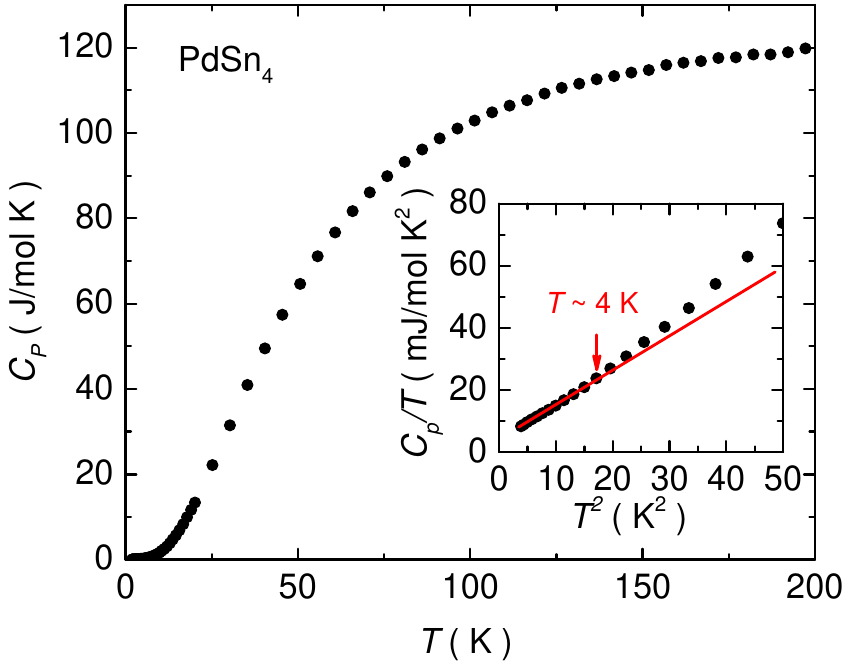}%
	\caption{(color online) Temperature dependent specific heat, $C_{P}$, of PdSn$_4$. Inset shows $C_{p}/T$ as a function of $T^2$. The red line is a linear fit to the lowest temperature PdSn$_4$ data. 
		\label{Speci}}
\end{figure}

The temperature dependent specific heat, $C_{P}$, of PdSn$_4$, shown in Fig.\,\ref{Speci}, does not show any signature of a phase transition. We plot the $C_{P}/T$ as a function of $T^{2}$ in the inset to Fig.\,\ref{Speci} to infer the electronic specific heat coefficient, $\gamma$, and the Debye temperature, $\Theta_{D}$. This is based on the relation $C_{P}\,=\,\gamma T + \beta T^{3}$ at low temperature; $\gamma\,=\,4$\,mJ/mol\,K$^2$ and $\Theta_{D}\,\sim\,206$\,K. These values are very similar to those of PtSn$_4$ ($\gamma\,\sim\,4$\,mJ/mol\,K$^2$ and $\Theta_{D}\,\sim\,210$\,K).\cite{Mun2012} Since both systems have $\gamma$ values of less than 1\,mJ/mol\,K$^2$\,atom, neither has a significant density of states at the Fermi surface or strong electron correlations.

\subsection{\texorpdfstring{Resistivity}{space}}

\begin{figure}
	\includegraphics[scale=1]{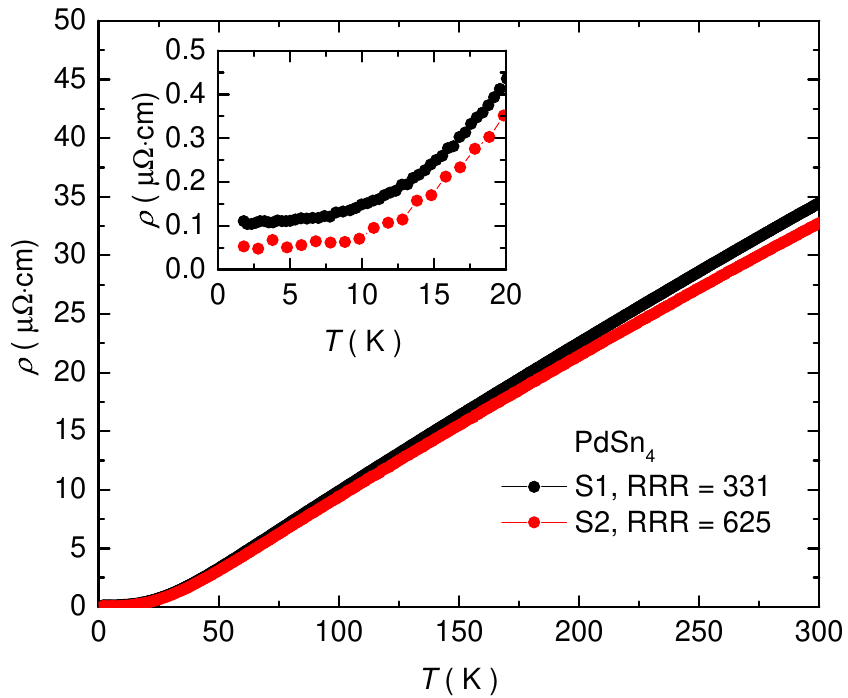}%
	\caption{(color online) Temperature dependent resistivity of two PdSn$_4$ single crystals. 
		\label{RT}}
\end{figure}

Figure\,\ref{RT} shows the temperature dependent resistivity, $\rho(T)$, measured from 1.8\,K to 300\,K. The current was applied perpendicular to the crystallographic $b$ direction. (i.e. within the plane of the plate-like sample). Sample S1 has a residual resistivity ratio (RRR\,=\,$\rho(300\,\textrm{ K})/\rho(1.8\,\textrm{K})$) value of 331, and sample S2 has RRR\,=\,625. In both cases, the large RRR indicates the high quality of the crystals. The overall behavior of $\rho(T)$ is metallic, which is similar to PtSn$_4$.\cite{Mun2012} It shows almost linear temperature dependence at high temperatures; the low temperature ($T\,<\,10$\,K) data from S1 and S2 were fitted with a polynomial function, $\rho\,(\textrm{T})\,=\,\rho_{0} + AT^{2}$. The obtained values are $\rho_{0}\,=\,0.105\,\mu\Omega\,\textrm{cm}$ and $A\,=\,2.7\,\times\,10^{-4}\,\mu\Omega\,\textrm{cm}/\textrm{K}^{2}$ for S1, and $\rho_{0}\,=\,0.049\,\mu\Omega\,\textrm{cm}$ and $A\,=\,2.2\,\times\,10^{-4}\,\mu\Omega\,\textrm{cm}/\textrm{K}^{2}$ for S2. Given that S2 has RRR, $\rho_{0}$ and $A$ values closer to those reported for PtSn$_4$ than S1,\cite{Mun2012} we chose to use S2 to carry out further transport experiments with the applied magnetic field. 

\begin{figure}
	\includegraphics[scale=1]{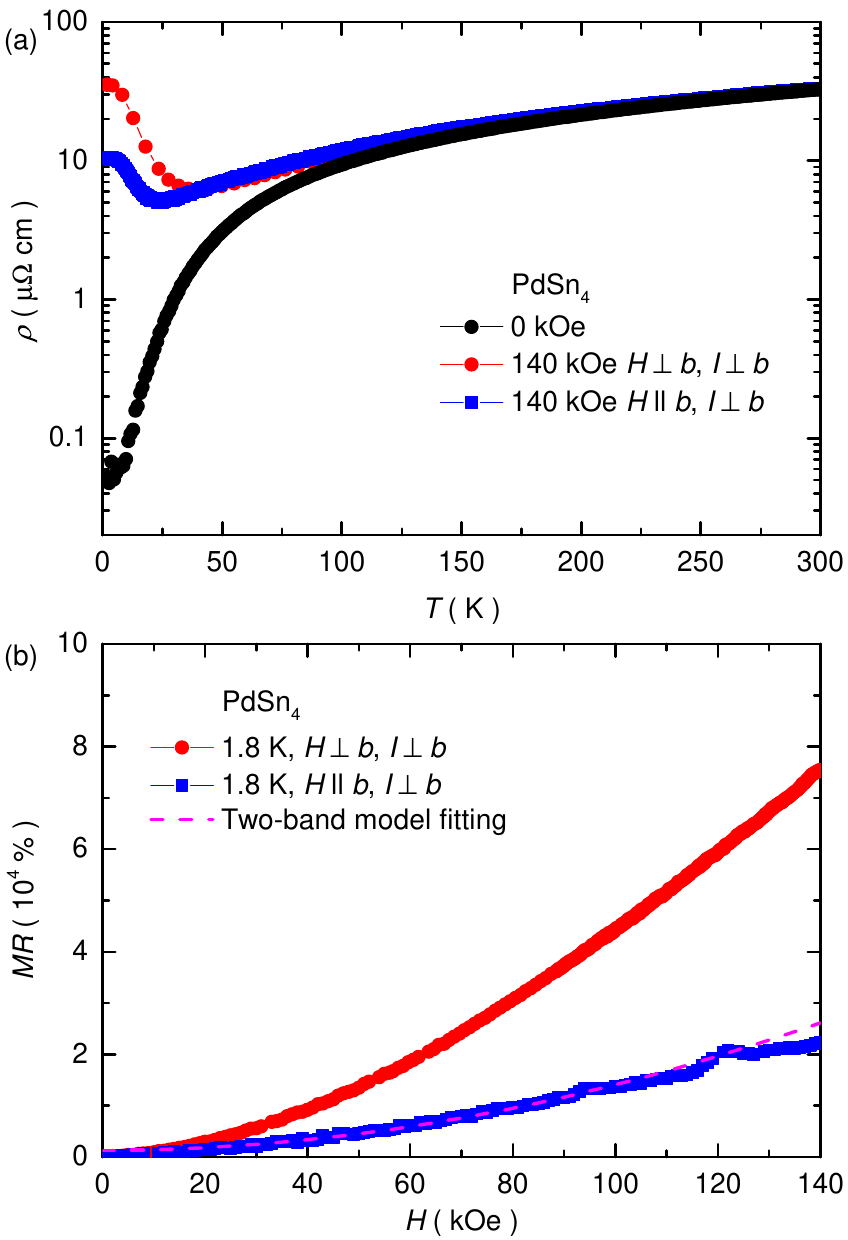}%
	\caption{(color online) (a) Temperature dependent resistivity of PdSn$_4$ in zero applied magnetic field (black filled circles), and in 140\,kOe for different magnetic field directions; red filled circles ($H\,\perp\,b$, $H\,\perp\,I$) and blue filled squares ($H\,\parallel\,b$, $H\,\perp\,I$). (b) Magnetoresistance (MR) of PdSn$_4$ at 1.8\,K for different magnetic field and current directions; red filled circles ($H\,\perp\,b$, $H\,\perp\,I$), blue filled squares ($H\,\parallel\,b$, $H\,\perp\,I$). Magenta dashed-line is fitted line based on a two-band model (see text). 
		\label{Electric}}
\end{figure}

The $\rho (T)$ of PdSn$_4$ for $H\,=\,140$\,kOe was measured from 1.8\,K to 300\,K with the two different magnetic field directions (red filled circles for $H\,\perp\,b$ and blue filled squares for $H\,\parallel\,b$) (Fig.\,\ref{Electric}\,(a)). Note that the current was always applied in an $I\,\perp\,b$ and $I\,\perp\,H$ direction. The current thus always flows in the $a$-$c$ plane, perpendicular to the applied magnetic field. For comparison, we also plotted the $\rho (T)$ of PdSn$_4$ for $H\,=\,0$\,kOe (black filled circles). For both field directions, PdSn$_4$ shows a strong field dependence at low temperatures. To be more specific, a local minimum is observed near 40\,K ($H\,\perp\,b$) and 25\,K ($H\,\parallel\,b$), followed by steep increases in resistivity that tends to saturate as $T\,\rightarrow\,2$\,K. 

Transverse magnetoresistance (MR) measurements were carried out from 0\,Oe to 140\,kOe at 1.8\,K, with MR defined as $(\rho(H)-\rho(H=0))\times\,100/\rho(H=0)$. The magnetic field dependency of MR in both directions is quadratic rather than linear, and there is no evidence of saturation in MR up to 140\,kOe. Anomalies in the higher magnetic field regime are due to quantum oscillations. (see discussion below). The order of the MR at 140\,kOe is 10$^4\,\%$ which is extremely large.   

As suggested by Fig.\,\ref{Electric} (a), the MR shown in Fig.\,\ref{Electric} (b) shows a clear anisotropy between the two different field directions: at 140\,kOe, MR = 7.5\,$\times\,10^{4}$\,$\%$ for $H\,\perp\,b$ and MR = 2.4\,$\times\,10^{4}$\,$\%$ for $H\,\parallel\,b$. It is interesting that both PtSn$_4$ and PdSn$_4$ show a larger MR when the magnetic field is applied perpendicular to the $b$ direction (i.e. within the basal plane). This is in contrast to many layered materials with large MR  anisotropy that manifest the largest MR when the magnetic field is applied perpendicular to the plane.\cite{Zhao2015, Eto2010} Furthermore, for layered materials that manifest quasi-two dimensional Fermi surfaces, it has been predicted that the transverse MR along one magnetic field direction (perpendicular to the plane) increases quadratically while it tends to saturate for a field that lies within the plane.\cite{lifshits1973} Given the quadratic MR, for both field directions, found for PdSn$_4$ as well as PtSn$_4$, it seems that the magetotransport is dominated by 3D Fermi surface (FS) sheets, even though each compound cleaves and exfoliates well. 
 
\subsection{\texorpdfstring{Hall coefficient}{space}}

\begin{figure}
	\includegraphics[scale=1]{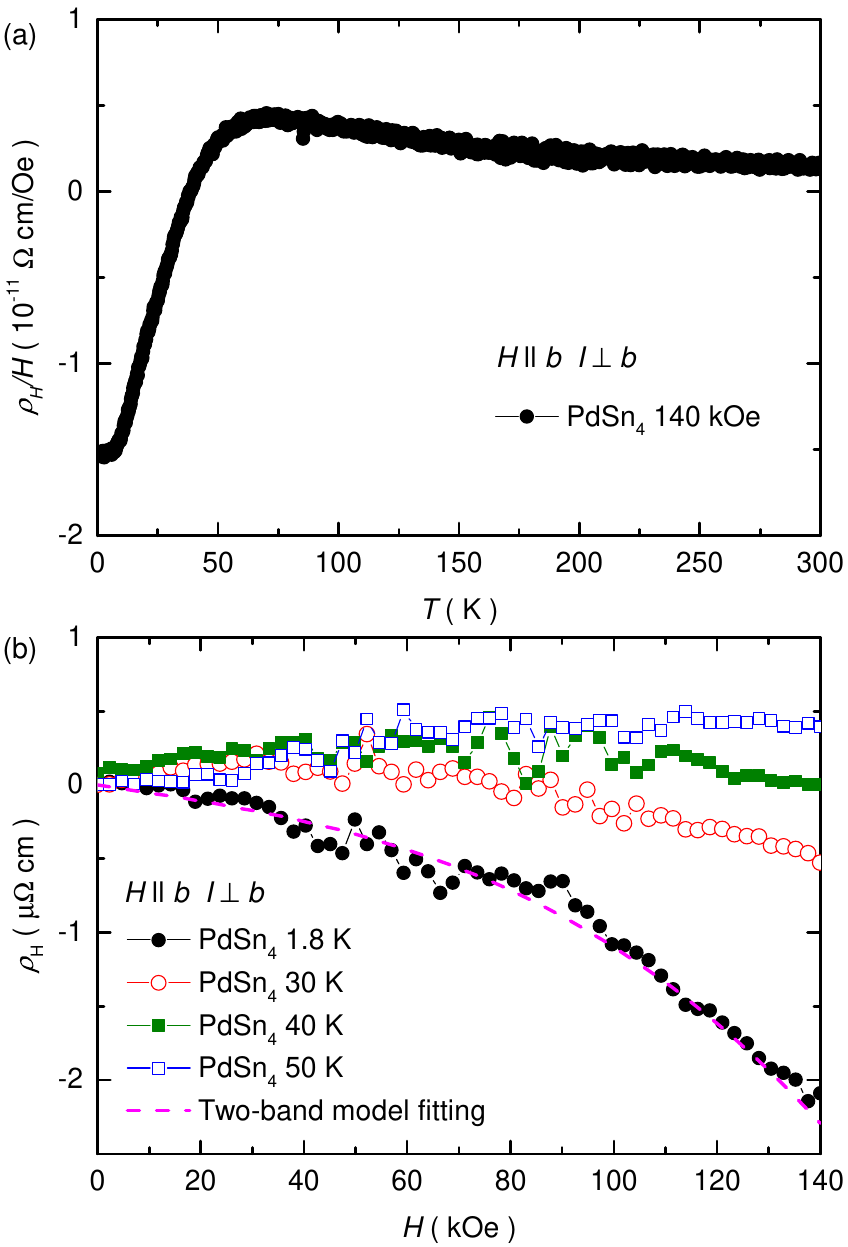}%
	\caption{(color online) (a) Temperature dependent Hall resistivity divided by the applied magnetic field, $\rho(H)/H$, of PdSn$_4$ for $H\,=\,140\,kOe$\,$\parallel\,b$. (b) magnetic field dependent Hall resistivity of PdSn$_4$ at 1.8\,K, 30\,K, 40\,K, and 50\,K. The dashed magenta line is based on a two-band model (see text).    
		\label{Hall}}
\end{figure}

Electron-hole compensation can be a reason for large MR and has been invoked in the case of WTe$_2$.\cite{Ali2014} In general, the conductivity tensor components $\sigma_{xy}$ and $\sigma_{yx}$ are proportional to 1/$H$ when the carrier density of electrons and holes are not approximately equal (i.e. $n_{e}\,\gg\,n_{h}$ or $n_{e}\,\ll\,n_{h}$). Then the diagonal components of resistivity tensors, except $\rho_{zz}$, saturate as the magnetic field goes to infinity. On the other hand, the resistivity tensor components $\rho_{xx}$ and $\rho_{yy}$ increase quadratically without saturating at high field when $n_{e}\,\approx\,n_{h}$ since $\sigma_{xy}$ and $\sigma_{yx}$ are then dominated by higher order of $1/H$ terms. One should note that this is true only if the following assumptions are fulfilled: (i) there are closed trajectories and (ii) in the high field regime.\cite{lifshits1973} In order to see whether such a compensation senario is possible for PdSn$_4$, we measured its Hall resistivity, $\rho_{H}$. In Fig.\,\ref{Hall} (a), the temperature-dependent Hall coefficient, $R_{H}\,=\,\rho_{H}/H$, of PdSn$_4$ at 140 kOe from 1.8\,K to 300\,K is plotted. Above 65\,K, $R_{H}$ of PdSn$_4$ shows monotonic behavior with a positive value which indicates that hole-like carriers dominate its transport properties. $R_{H}$ of PdSn$_4$ starts to decrease rapidly below 65\,K, crosses $R_{H}\,=\,0$ at 38\,K, and then saturates to $R_{H}\,\approx\,-1.5\,\times\,10^{-11}$\,\ohm\,cm/Oe below 7\,K. Given that $R_{H}$ changes sign at low temperatures, it is tempting to associate this phenomenon with perfect carrier compensation. However, this assumes that the mobilities of all carrieres are equal. This is because $R_{H}$ is not only a function of carrier density but also mobility. Therefore, we need a way to disentangle the carrier density and mobility contributions to $R_{H}$.

One indication of compensation is a nonlinear behavior of $\rho_{xy}(H)\,\equiv\,\rho_{H}$ in high magnetic fields.\cite{lifshits1973} Thus, we measured the $\rho_{H}$ of PdSn$_4$ as a function of the magnetic field in Fig.\,\ref{Hall} (b). The $\rho_{H}$ of PdSn$_4$ shows nonlinear behavior in the low temperature regime, where it shows non-saturating, large MR. This means PdSn$_4$ has very similar values of carrier densities, i.e. $n_{e}\,\approx\,n_{h}$, whereas the mobility of electrons is larger compared to the mobility of holes based on $R_{H}(T)$ at low temperature (see Fig.\,\ref{Hall} (a)).    

We were able to use a simplified two-band model to fit the $\rho_{H}(H)$ and MR data at 1.8\,K in order to do a quantitative analysis.\cite{Pippard1989} This method has been successfully used in many materials.\cite{Rullier2009, Rullier2010, Luo2015}

\begin{equation}
{ \rho  }_{ xx }\left( B \right) =\frac { 1 }{ e } \frac { \left( { n }_{ h }{ \mu  }_{ h }+{ n }_{ e }{ \mu  }_{ e } \right) +\left( { n }_{ h }{ \mu  }_{ e }+{ n }_{ e }{ \mu  }_{ h } \right) { \mu  }_{ h }{ \mu  }_{ e }{ B }^{ 2 } }{ { \left( { n }_{ h }{ \mu  }_{ h }+{ n }_{ e }{ \mu  }_{ e } \right)  }^{ 2 }+{ \left( { n }_{ h }{ -n }_{ e } \right)  }^{ 2 }{ { \mu  }_{ h } }^{ 2 }{ { \mu  }_{ e } }^{ 2 }{ B }^{ 2 } } 
\end{equation}
 
\begin{equation}
{ \rho  }_{ xy }\left( B \right) =\frac { B }{ e } \frac { \left( { n }_{ h }{ { \mu  }_{ h } }^{ 2 }{ -n }{ _{ e }{ \mu  }_{ e } }^{ 2 } \right) +\left( { n }_{ h }-{ n }_{ e } \right) { { \mu  }_{ h } }^{ 2 }{ { \mu  }_{ e } }^{ 2 }{ B }^{ 2 } }{ { \left( { n }_{ h }{ \mu  }_{ h }+{ n }_{ e }{ \mu  }_{ e } \right)  }^{ 2 }+{ \left( { n }_{ h }{ -n }_{ e } \right)  }^{ 2 }{ { \mu  }_{ h } }^{ 2 }{ { \mu  }_{ e } }^{ 2 }{ B }^{ 2 } }  
\end{equation} 

The values obtained from the two-band model are $ n_e = 1.6 \times 10^{27} \text{m}^{-3} \pm 0.1 \times 10^{27} \text{m}^{-3}$, $n_h = 1.5 \times 10^{27} \text{m}^{-3} \pm 0.1 \times 10^{27} \text{m}^{-3}$ such that their ratio is $c = n_h/n_e = 0.93 \pm 0.02$. The mobilities are found to be $\mu_e = 0.34 \pm 0.04$\,m$^{2}$/V\,s and $\mu_h = 0.27 \pm 0.03$\,m$^{2}$/V\,s such that their ratio is $c_\mu = \mu_h/\mu_e = 0.80 \pm 0.05$. The corresponding $R^2$-values of the fit are very good $R^2 - 1 = 0.02$, showing that although the Fermi surface clearly consists of several hole and electron pockets with different mobilities and effective masses (see more below), the resistivity can be well captured (at least qualitatively) within a two-band model. We want to emphasize that the simplified two-band model equations above are based on the assumption of (i) only one electron and one hole band are present, (ii) the isotropic free electron gas model is applicable, and  (iii) components of the conductivity tensor that contain $z$ need not be considered. In addition, we have found that slightly worse but still reasonable fits can be obtained with other mobility ratios $c_\mu$ and total densities $n_e$. The data constrains the ratio $c$ more than the value of $n_e$.\cite{commentary}

The best fits we have obtained are plotted in both Fig.\,\ref{Electric} (b) and Fig.\,\ref{Hall} (b) with magenta dashed lines. Unlike PtSn$_4$,\cite{Mun2012} the values of $n_{e}$ and $n_{h}$ for PdSn$_4$ obtained from the fit are very similar, within about 7\,$\%$ of each other, whereas $\mu_{e}\,>\,\mu_{h}$, which was expected from the Hall data at the high magnetic fields. The shape of $\rho_{xy}$ immediately yields two constraints $(i) n_e \mu_e^2 > n_h \mu_h^2$ and $n_h < n_e$, which follows directly from the fact that the linear and cubic coefficients in an expansion $\rho_{xy} = a_1 B + a_3 B^3$ both need to be negative $a_1,a_3 < 0$. From the fact that the non-linearity sets in around $B_2 \approx 8$T, we can further derive the relation $(1-c c_\mu^2)(1 + c c_\mu)^2/[c c_\mu^2 (1-c) (1+c_\mu)^2] \approx B_2^2 \mu_e^2$.

\subsection{\texorpdfstring{Magnetization}{space}}

\begin{figure}
	\includegraphics[scale=1]{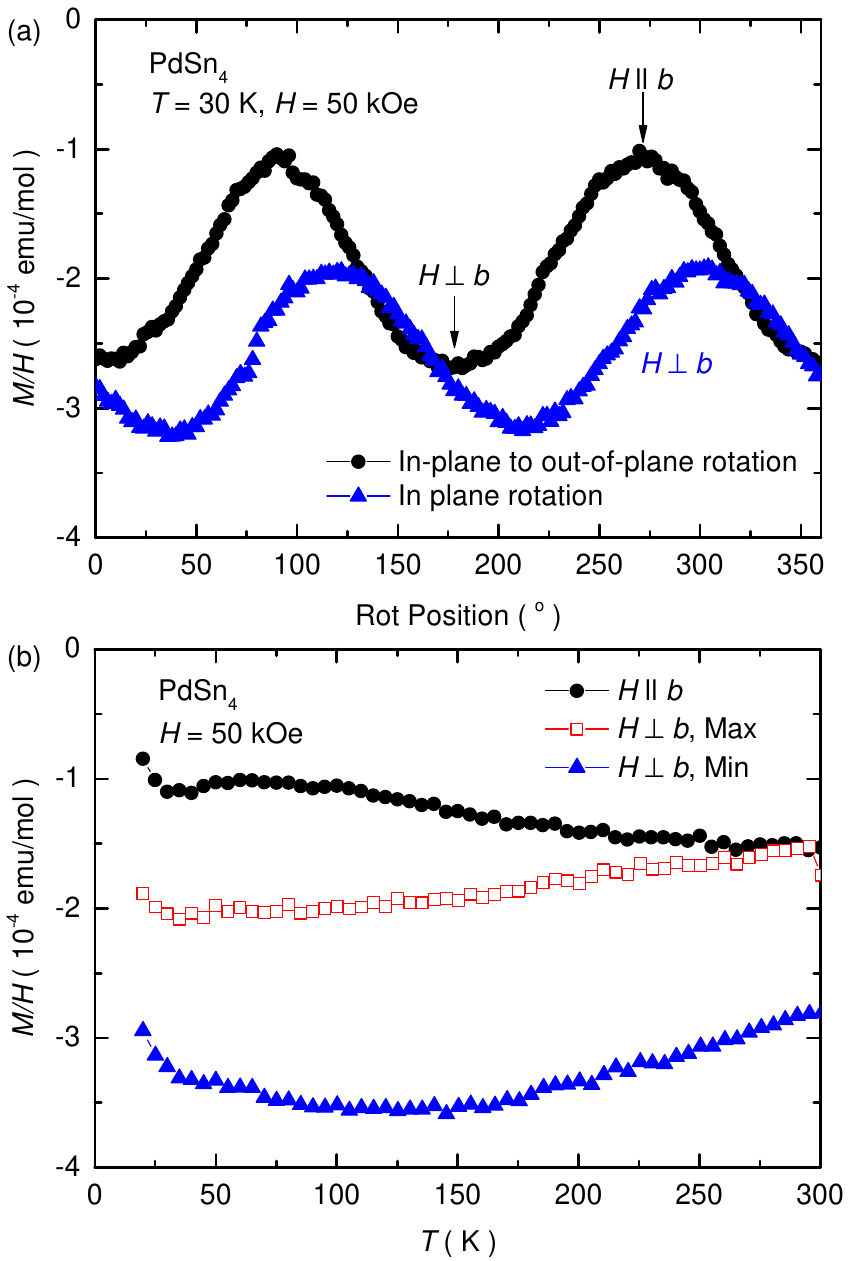}%
	\caption{(color online) (a) Angular-dependent magnetic susceptibility of PdSn$_4$ in both in-plane to out-of-plane rotation and in-plane rotation at $T\,=\,30$\,K and $H\,=\,50$\,kOe. (b) Temperature-dependent magnetic susceptibility of PdSn$_4$ at 50\,kOe, $H\,\parallel\,b$ (black filled circles), $H\,\perp\,b$ for maximal in-plane magnetization orientation (red open rectangles) and $H\,\perp\,b$ for minimal in-plane magnetization orientation (blue filled triangles). 
		\label{Magnetic}}
\end{figure}

Both out-of-plane and in-plane angular dependent magnetization measurements are shown in Fig.\,\ref{Magnetic} (a). The observed magnetic properties of PdSn$_4$ are diamagnetic in all directions with a clear anisotropy between the $a$-$c$ plane and the $b$ direction. The data are taken at 30\,K and 50\,kOe, but the magnetization is linear as a function of the magnetic field for $H\,\le\,7$\,T, so the angular dependent magnetic susceptibilities measured in different magnetic fields of 25\,kOe, 50\,kOe and 70\,kOe are the same. On the other hand, both angular dependent magnetic susceptibility show deviations from simple sine function behavior when the temperature is below 30\,K due to quantum oscillations.

Based on the angular dependent magnetization measurements, we find that the maximum susceptibility (smallest negative value) occurs for $H\,\parallel\,b$. For $H\,\perp\,b$, we find a simple angular variation of $M(\theta)$ and can identifiy an in-plane field orientation that produces a minimal and maximal response separated by 90$^{o}$. In Fig.\,\ref{Magnetic} (b) we present the temperature dependence magnetization from 20\,K to 300\,K with 50 kOe along the three different magnetic field directions. All of them show weak temperature dependence, and the anisotropy gets smaller as temperature is increased. The low temperature upturn is due to entering the temperature range of quantum oscillations (see below).

In order to understand the physical meaning of an anisotropic susceptibility, it is important to separate core diamagnetism from Pauli paramagnetism and Landau diamagnetism. If we subtract the core diamagnetic contribution,\cite{boudreaux1976,BainBerry2008} we can infer the conduction electron contribution to the magnetic susceptibility. Even after the core diamagnetic correction, $\chi_{core}\,=\,- 1.05\,\times\,10^{-4}$\,emu/mol of PdSn$_4$,\cite{boudreaux1976,BainBerry2008} the remaining conduction electron susceptibility is still essentially diamagnetic. Based on an equation for the total electronic susceptibility of a metal\cite{Blundell2001}
\begin{equation}
\chi ={ \chi  }_{ P }\left[ 1-\frac { 1 }{ 3 } { \left( \frac { { m }_{ e } }{ { m }^{ * } }  \right)  }^{ 2 } \right], 
\end{equation}
where $m_{e}$ is an electron mass, m$^{*}$ is an effective mass and $\chi_{P}$ is Pauli paramagetism,
the result indicates that the carriers should have rather low effective masses.

\begin{figure}
	\includegraphics[scale=1]{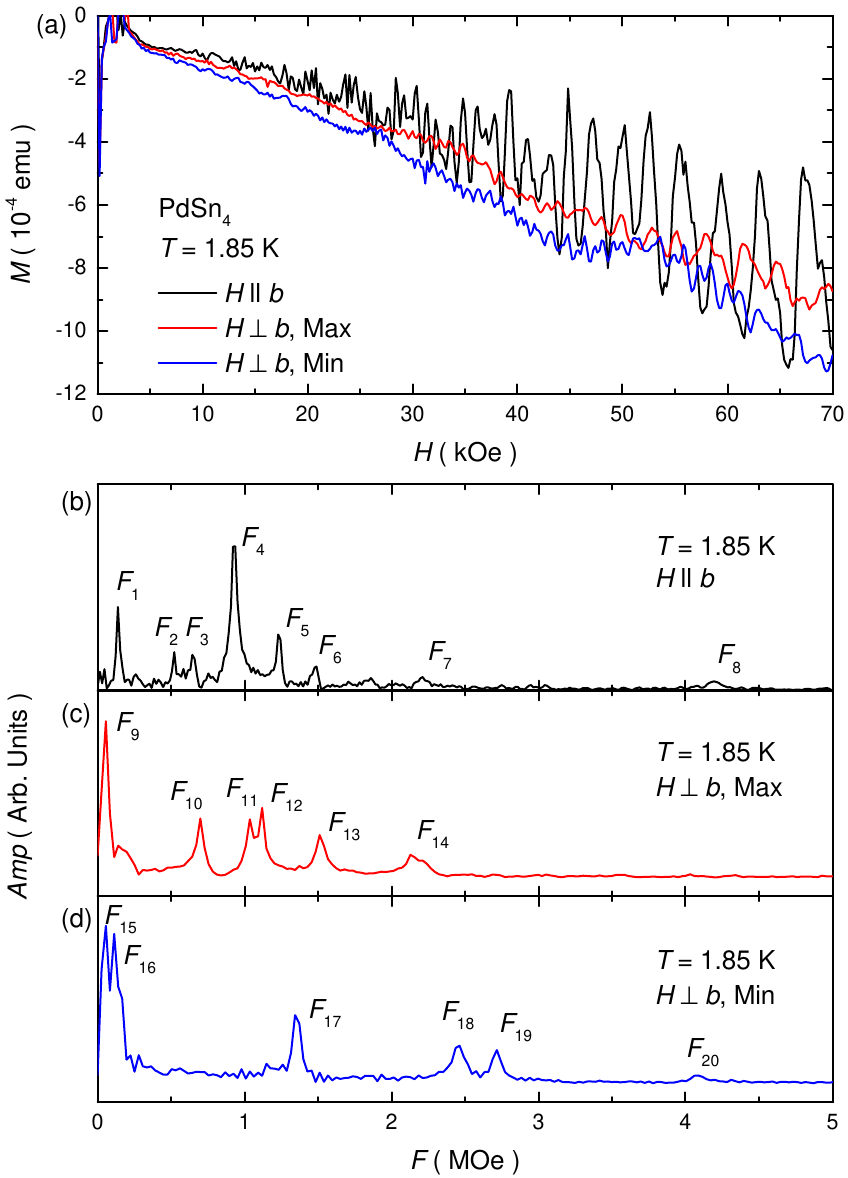}%
	\caption{(color online) (a) Magnetization isotherms of PdSn$_4$ without background subtraction for $H\,\parallel\,b$, $H\,\perp\,b$, Max and $H\,\perp\,b$, Min at 1.85\,K. (b) FFT spectra of dHvA data for $H\,\parallel\,b$. (c) FFT spectra of dHvA data for $H\,\perp\,b$, Max. (d) FFT spectra of dHvA data for $H\,\perp\,b$, Min.
		\label{QOdirection}}
\end{figure}

More detailed information about the Fermi surface and effective masses can be obtained via measurement and analysis of quantum oscillations. Thus, we measured the magnetization of PdSn$_4$ as a function of the magnetic field, $M(H)$, up to 70 kOe at 1.85 K in all three salient directions: $H\,\parallel\,b$, $H\,\perp\,b$, Max and $H\,\perp\,b$, Min as shown in Fig.\,\ref{QOdirection}\,(a). Note that the plotted magnetization data are raw data without subtracting the background of the sample rotator, but clear de Haas van Alphen (dHvA) oscillations are readily detected. For analysis, the oscillatory part of the magnetization is obtained as a function of $1/B$ after subtracting a linear background. We then used a fast fourier transform (FFT) algorithm. The FFT data are shown in Fig.\,\ref{QOdirection}\,(b) for $H\,\parallel\,b$, Fig.\,\ref{QOdirection}\,(c) for $H\,\perp\,b$, Max and Fig.\,\ref{QOdirection}\,(d) for $H\,\perp\,b$, Min. All the frequencies are labeled on the figures and listed in the Table\,\ref{table1}. $F_\textrm{14}$ is broad, and it might be a superposition of the second harmonics of $F_\textrm{11}$ and $F_\textrm{12}$. In addition, $F_\textrm{19}$ is likely a second harmonic of $F_\textrm{17}$. Here, we note that Shubnikov-de Haas oscillation show some low frequencies ($F_{1}$, $F_{2}$ and $F_{3}$) for $H\,\parallel\,b$ as noted in Fig.\,\ref{Electric} (b), but these oscillations are much clearer in the $M(H)$ data. The corresponding extremal areas were calculated via the the Onsager relation which gives a direct proportionality between frequency and extremal area, $F_\textrm{i} = \frac{\hbar c}{2 \pi e}S_{i}$. 

\begin{table*}[]
	\centering
	\caption{Frequencies of the peaks from Fig.\,\ref{QOdirection} (b)-(d) and Fig.\,\ref{Oscillation} (c), corresponding orbital areas $S_{i}\,=\,2\,\pi\,e\,F_{i}/(\hbar\,c)$, and effective masses from Fig.\,\ref{Oscillation}}
	\label{table1}
	\begin{tabular}{lllllllllllllllllllllllllll}
		\hline \hline
		\multicolumn{8}{c}{} & \multicolumn{11}{c}{Rotator} & & \multicolumn{7}{c}{Plastic disk}\\
		\multicolumn{7}{c}{$H\,\parallel\,b$} & &  \multicolumn{5}{c}{$H\,\perp\,b$, Max} & & \multicolumn{5}{c}{$H\,\perp\,b$, Min} & & \multicolumn{7}{c}{$H\,\perp\,b$}\\
		        & & MOe & & $\AA^{-2}$ & & m$_e$ & & & & MOe & & $\AA^{-2}$ & & & & MOe & & $\AA^{-2}$ & & & & MOe & & $\AA^{-2}$ & & m$_e$ \\\hline
		$F_{1}$ & & 0.14 & & 0.0013 & & & & $F_{9}$ & & 0.06 & & 0.00057 & & $F_{15}$ & & 0.06 & & 0.00057 & & $F^{*}_{9}$ & & 0.06 & & 0.00057 & & 0.03 \\
		$F_{2}$ & & 0.52 & & 0.0050 & & 0.07 & & $F_{10}$ & & 0.70 & & 0.0067 & &$F_{16}$ & & 0.11 & & 0.0011 & & $F^{*}_{10}$ & & 0.76 & & 0.0073 & & 0.05  \\
		$F_{3}$ & & 0.65 & & 0.0062 & & 0.08 & & $F_{11}$ & & 1.04 & & 0.0099 & & $F_{17}$ & & 1.34 & & 0.0128 & & $F^{*}_{11}$ & & 1.06 & & 0.0101 & & 0.05    \\ 
		$F_{4}$ & & 0.94 & & 0.0090 & & 0.09 & & $F_{12}$ & & 1.12 & & 0.0107 & & $F_{18}$ & & 2.46  & & 0.0235 & & & & & & & &  \\
		$F_{5}$ & & 1.23 & & 0.0118 & & 0.08 & & $F_{13}$ & & 1.51 & & 0.0144 & & $F_{19}$ & & 2.72 & & 0.0260 & & $F^{*}_{13}$ & & 1.57 & & 0.015 & & 0.05     \\
		$F_{6}$ & & 1.47 & & 0.0140 & & 0.08 & & $F_{14}$ & & 2.13 & & 0.0203 & & $F_{20}$ & & 4.09 & & 0.0391 & & $F^{*}_{14}$ & & 2.13 & & 0.0203 & & 0.06     \\
		$F_{7}$ & & 2.20 & & 0.0210 & & 0.104 & & & & & & & & & & & & & & & & & & & & \\
		$F_{8}$ & & 4.18 & & 0.0399 & & 0.106 & & & & & & & & & & & & & & & & & & & & \\ \hline \hline	        
	\end{tabular}
\end{table*}

\begin{figure*}
	\includegraphics[scale=0.8]{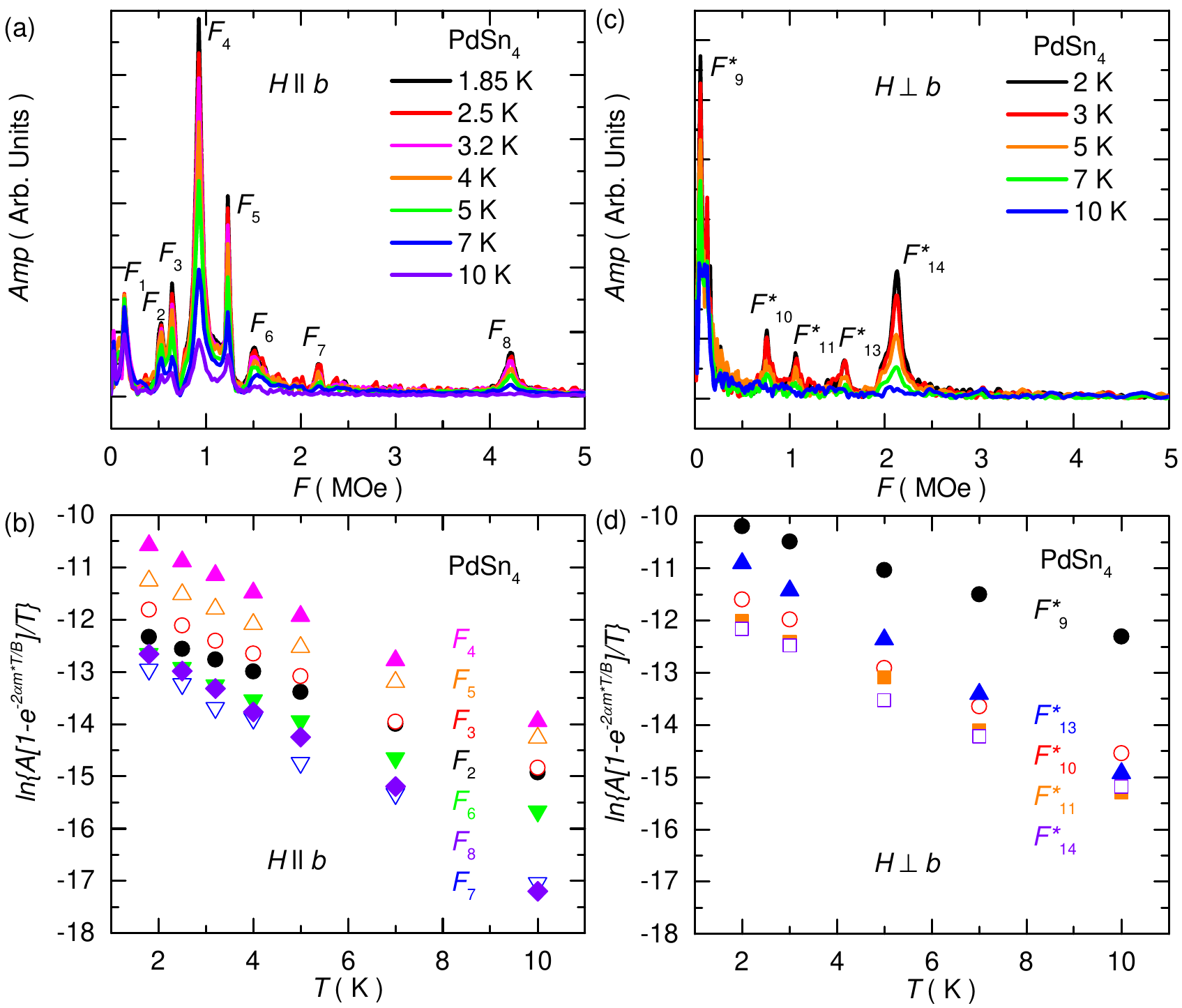}%
	\caption{(color online) (a) Temperature dependence of the fast Fourier transformed dHvA data of PdSn$_4$ for $H\,\parallel\,b$ and $T$\,=\,1.85\,K, 2.5\,K, 3.2\,K, 4\,K, 5\,K, 7\,K, 10\,K. (b) Mass plot of dHvA orbits at each frequency when the magnetic field was applied parallel to the crystallographic $b$-axis. (c) Temperature dependence of the fast Fourier transformed dHvA data of PdSn$_4$ for $H\,\perp\,b$ and $T$\,=\,2\,K, 3\,K, 5\,K, 7\,K, 10\,K. (d) Mass plot of dHvA orbits at each frequency when the magnetic field was applied perpendicular to the crystallographic $b$-axis. 
		\label{Oscillation}}
\end{figure*}

$M(H)$ data were measured not only at the base temperature, 1.85\,K, but also at the several other temperatures, 2.5 K, 3.2 K, 4 K, 5 K, 7 K and 10 K. (Fig.\,\ref{Oscillation}) For the measurement of these magnetic isotherms, the sample was mounted on a diamagnetic, isotropic, amorphous disk and manually rotated with respect to the applied field. Exactly the same peaks that we detected from  rotator (Fig.\,\ref{QOdirection} (b)) are found when the $H\,\parallel\,b$. On the other hand, the FFT data for $H\,\perp\,b$ show slightly different peaks compared to both $H\,\perp\,b$, Max and $H\,\perp\,b$, Min measured from the rotator, although the frequency of the peaks are very similar to that of $H\,\perp\,b$, Max, except for $F_{12}$.

Figure.\,\ref{Oscillation} shows that the oscillation amplitudes decrease as temperature increases. The Lifshitz-Kosevich formula\cite{Shoenberg1984} explains a phase smearing which is related to amplitude. Specifically, the effect of temperature is described by
\begin{equation}
{ R }_{ T }=\frac { \alpha { m }^{ * }T/H }{ \sinh(\alpha { m }^{ * }T/H) } 
\end{equation} 
where $\alpha=2\pi^{2} c k_{B}/e \hbar $. We used an average value for $1/H\,=\,(1/H_{min}+1/H_{max})*1/2$, where $H_{min}$ and $H_{max}$ indicate the range of the data, we used for FFT. Effective masses are obtained based on the above equation, and linear behavior in mass plot in Fig.\,\ref{Oscillation} (b) and (d) suggests that the obtained effective masses are reasonable. The effective mass of each frequency is shown in the Table.\,\ref{table1}. All effective masses are small and approximately equal to each other. However, the low frequency peak, $F^{*}_{9}$ has a particularly small effective mass which suggests a potentially large contribution to density of states. Thus, the carriers confined to the Fermi surface related to the $F^{*}_{9}$ may give a dominant contribution to the anisotropy of the magnetic susceptibility.  

\section{Discussion and comparison to P\lowercase{t}S\lowercase{n}$_4$}

The temperature and field dependent properties, including anisotropies, of PdSn$_4$ are qualitatively similar to those of PtSn$_4$,\cite{Mun2012} but there exist some important quantitative differences. The question that we now would like to address is whether a comparison of these two, related, materials help us understand the origin of XMR.  

\subsection{\texorpdfstring{Carrier compensation}{space}}

Carrier compensation is one of the reasons that many materials, including WTe$_2$,\cite{Ali2014} LaBi,\cite{tafti2016} LaSb\cite{tafti2016} and PtBi$_2$\cite{Gao2017}, are thought to exhibit XMR behavior. However, the scenario of nearly perfect compensation of carriers cannot explain XMR in  PdSn$_4$ and PtSn$_4$. At the simplest level, PdSn$_4$ exhibits a larger degree of compensation than PtSn$_4$ but shows a smaller, MR. At low temperatures where MR is large, PdSn$_4$ shows a non-linear field dependence of $\rho_{H}(H)$ at high field whereas a linear $\rho_{H}$ is observed for PtSn$_4$ in the high field regime.\cite{Mun2012} Note that such non-linear behavior is an indication of carrrier compensation. Moreover, the results from classical two-band-model fitting indicate that PdSn$_4$ has similar electron and hole carrier densities albeit with different mobilities, whereas PtSn$_4$ has an almost two order of magnitude difference between electron and hole carrier densities.\cite{Mun2012} (It should be noted, though, that if we consider classical carrier compensation theory, this describes the non-saturating $H^2$ behavior but not the prefactor.)\cite{lifshits1973} This means that, in this simplest treatment, the magnitude of MR cannot be explained. For the more specific case of WTe$_2$ an exitonic insulator was invoked;\cite{Ali2014} for PdSn$_4$ and PtSn$_4$ such an insulating state is unlikely given their multiple and complex Fermi surfaces. Finally, a recent theoretical paper studied the compensated Hall effect in confined geometry with electron-hole recombination near the edge, inducing a large MR.\cite{Alekseev2017} In this case, linear response regime is found when the edge currents dominate the resistance.\cite{Alekseev2017} On the other hand,  for PdSn$_4$ and PtSn$_4$ we find quadratic behavior up to 140\,kOe. We can thus confidently rule out (i) carrier compensation, (ii) the emergence of an excitonic insulating state, and (iii) MR enhancement by sample confinement as the primary origins of XMR in PdSn$_4$ and PtSn$_4$. 

\subsection{\texorpdfstring{Dirac arc node feature}{space}}

\begin{figure*}[tb]
	\includegraphics[width=6.5in]{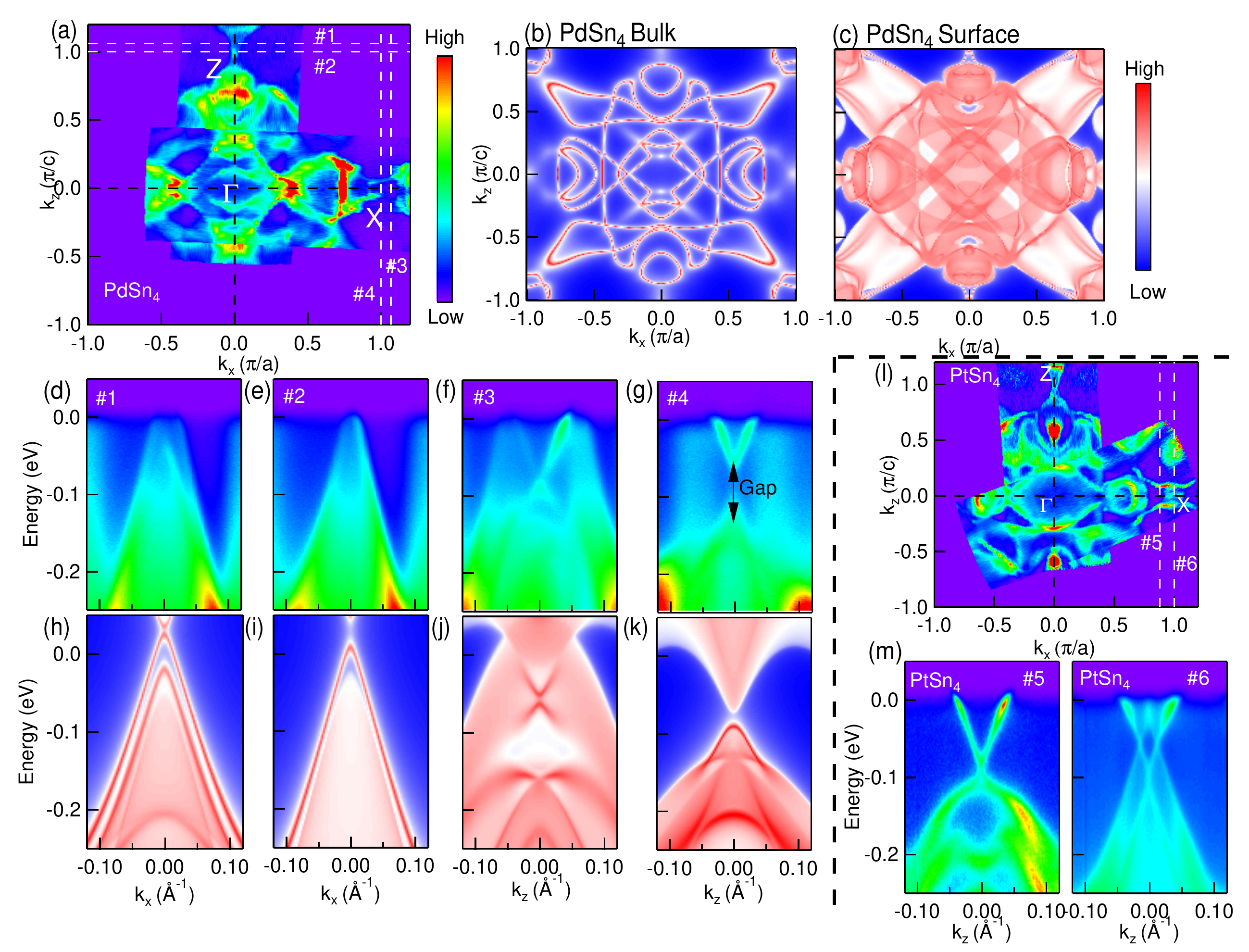}%
	\caption{Experimental ARPES and calculated results for the electronic Fermi surface and band dispersion of PdSn$_{4}$. 
		(a) Fermi-surface plot of ARPES intensity integrated within 10\,meV of the chemical potential along $\Gamma-Z$ and $\Gamma-X$.
		(b) Bulk FS calculated by density functional theory (DFT) at $k_{y} = 0.364~\pi/b$. 
		(c) Calculated surface FS.
		(d)-(g) ARPES band dispersions along cut \#1-\#4 in (a), respectively.
		(h)-(k) Calculated surface band dispersion corresponding to (d)-(g).
		(l) For comparison, Fermi-surface plot of ARPES intensity integrated within 10 meV of the chemical potential along $\Gamma-Z$ and $\Gamma-X$ in PtSn${}_{4}$.\cite{WuPtSn2016}
		(m) Band dispersion in PtSn${}_{4}$ at $k_x = 0.88 \text{ and } 1.0~\pi/a$ along cut \#5 and 6 in (l).\cite{WuPtSn2016}
		\label{fig:Fig1}}
\end{figure*}

\begin{figure*}[tb]
	\includegraphics[width=5in]{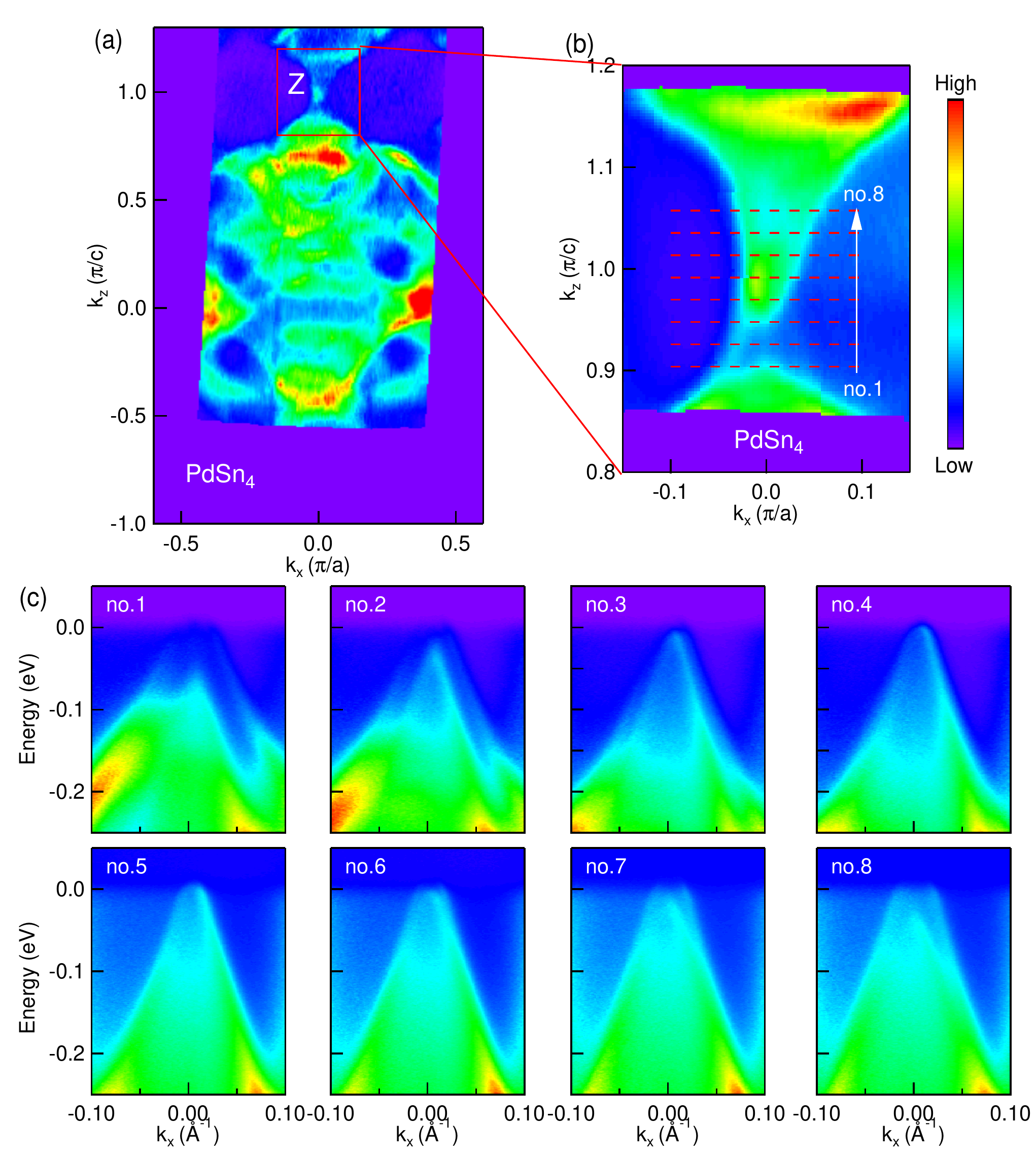}%
	\caption{Fermi surface and band dispersion in the proximity of the Z point,(0,1), of PdSn$_4$.
		(a) Fermi surface plot of the ARPES intensity integrated within 10~meV of the chemical potential along $\Gamma-Z$. 
		(b) Zoomed image of the red box in (a), red dashed lines mark cuts no.1 –- no.8. 
		(c) Band dispersion along cuts no. 1 –- no. 8. Cut no. 5 passes through the Z point. 
		\label{fig:Fig2}}
\end{figure*}

\begin{figure*}[tb]
	\includegraphics[width=6.5in]{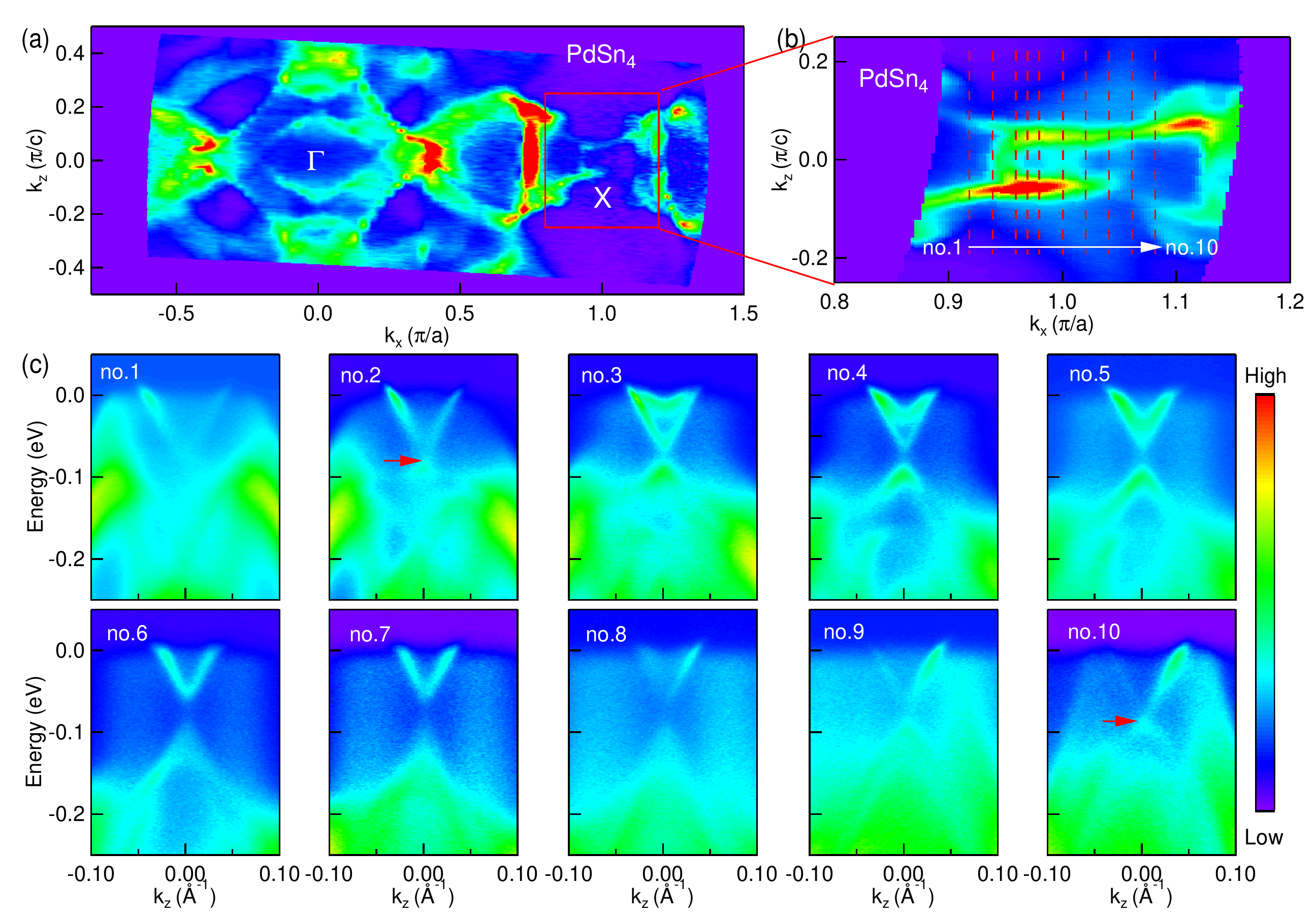}%
	\caption{Fermi surface plot and band dispersion close to the X point, (1,0), of PdSn$_4$.
		(a) Fermi surface plot of the ARPES intensity integrated within 10 meV of the chemical potential along $\Gamma-X$. 
		(b) Zoomed image of the red box in (a), red dashed lines mark cuts no.1 -- no.10. 
		(c) Band dispersion along cuts no.1 –- no.10. Cut no. 6 passes through the X point. 
		\label{fig:Fig3}}
\end{figure*}

In order to check whether the Dirac arc node feature seen in PtSn$_4$\cite{WuPtSn2016} can be associated with XMR in PdSn$_4$, or not, angle resolved photo emission spectroscopy (ARPES) measurements and surface band structure calculations were performed. The Fermi surface and band dispersion along key directions in the Brillouin Zone (BZ) for PdSn$_{4}$ are shown in Fig.~\ref{fig:Fig1}. Panel (a) shows the ARPES intensity integrated within 10 meV about the chemical potential. High intensity areas mark the contours of the FS sheets. The FS consists of at least two electron pockets at the center of BZ surrounded by several other electron and hole FS sheets. Fig.~\ref{fig:Fig1} (b) shows the calculated bulk FS, which matches the ARPES data well close to the center of the zone in Fig.\,\ref{fig:Fig1} (a). However, it does not predict the linear band dispersion near the $Z$ point (see below) or the FS crossings close to the $X$ point, missing a set of curved FS sheets that are present in Fig.\,\ref{fig:Fig1} (a). This is similar to the results of PtSn$_4$, where a surface state calculation was needed to reproduced the Fermi surface sheets in the proximity of the $X$ point.\cite{WuPtSn2016} Thus we carried out a surface spectral function calculation as shown in Fig.\,\ref{fig:Fig1} (c) and we can clearly see extra FS sheets connecting the bands close to the center of BZ and at the $X$ point. Band dispersion along several cuts in proximity of the $Z$ and $X$ points are shown in Figs.\,\ref{fig:Fig1} (d)--(g). Close to the $Z$ point (Figs.\,\ref{fig:Fig1} (d) and (e)), the dispersion resembles a Dirac-like feature. This linear dispersion was not found in bulk calculations and only appeared in the surface spectral function calculations (Figs.\,\ref{fig:Fig1} (h) and (i)). The small gap in the Dirac-like feature at the Z point (Figs.\,\ref{fig:Fig1} (e) and (i)) is reduced to zero for cuts further away from the BZ boundary (Figs.\,\ref{fig:Fig1} (d) and (h)). Close to the $X$ point [Fig.\,\ref{fig:Fig1} (f)], the band dispersion consists of Dirac-like dispersion (surface origin) surrounded by broad hole bands (most likely bulk states). The data agree well with surface spectral function calculation shown in Fig.\,\ref{fig:Fig1} (j). In Fig.\,\ref{fig:Fig1} (g), the observed conduction band and valence band have a significant band gap as marked by the black arrows, which is in agreement with the surface spectral function calculation shown in Fig.\,\ref{fig:Fig1} (k). 

The data in Figs.\,\ref{fig:Fig1} (a), (d)--(g) demonstrate that the experimentally observed band structure has both bulk and surface components, and the Dirac-like features near Z and X points have surface origin. Compared to the results of PtSn$_4$ shown in Fig.\,\ref{fig:Fig1} (l), the Fermi surface of these two compounds are very similar close to the zone center, which is consistent with the similar crystal structures and electronic configurations of the compounds. However, comparing the results in Figs.\,\ref{fig:Fig1} (f) and (g) with those in Figs.\,\ref{fig:Fig1} (m), the primary difference between these two data sets is that for PdSn$_{4}$,  the double Dirac node arcs surface state is gapped out [Fig.\,\ref{fig:Fig1} (g)]. 

In Fig.~\ref{fig:Fig2}, we show details of the features of PdSn$_4$ near the Z point. An enlarged image from the red box in Fig.~\ref{fig:Fig2}(a) is shown in panel (b), where two triangular-shaped FS sheets make a somewhat disconnected hourglass like shape. The detailed evolution of band dispersions along cuts no.~1 to no.~8 is shown in Fig.~\ref{fig:Fig2}(c). A sharp linear dispersion (the inner band) starts to cross at a binding energy of $\sim$100~meV in cut no.~1; this band moves up in energy and eventually becomes almost degenerate with the outer bands in cut no.~5. As we move further toward the second BZ cuts(no.~6 -- no.~8), the sharp band moves back down in energy, as expected.

A similar study of the band evolution near the X point is shown in Fig.~\ref{fig:Fig3}. We focus on the Fermi surface and band dispersion in a small area in the part of the BZ that is marked by the red box in Fig.~\ref{fig:Fig3}(a). The Fermi surface in this region consists of two parabolic FS sheets intersecting each other, forming two almost parallel straight segments in the overlapping region. Detailed band dispersion along cuts no.~1 to no.~10 are shown in Fig.~\ref{fig:Fig3}(c). The data along cut no.~1 show a sharp electron band enclosed by the bulk intensity. As we move to cut no.2, a Dirac-like dispersion is revealed with the possible Dirac point marked by the red arrow. Moving closer to the X point (cut no.~3), we can see a gap develops between the conduction and valence bands, and another electron band emerges. As we move to cuts no.~4, 5, and 6, the two electron bands come closer and eventually merge together in cut no.~6 at the X point. A significant gap between the conduction and valence bands persists across these cuts. The gap size slowly decreases and seems to form another Dirac dispersion in cut no.~10 in the second BZ. The observed features around X point are markedly different for PtSn$_{4}$,\cite{WuPtSn2016} where gapless Dirac node arc structure is observed. 

Even though, in PdSn$_4$ the Dirac point at $Z$ is a surface state and the Dirac arc node feature is gapped out, XMR behavior is still present in both compounds. Moreover, both PdSn$_4$ and PtSn$_4$ show larger MR when the magnetic field is applied perpendicular to the $b$ direction. This is not consistent with the measured surface states at a surface perpendicular to $b$ giving rise to the measured XMR, because the resulting orbits do not lie in the surface plane.

\subsection{\texorpdfstring{Kohler's rule}{space}}

\begin{figure}
	\includegraphics[scale=1]{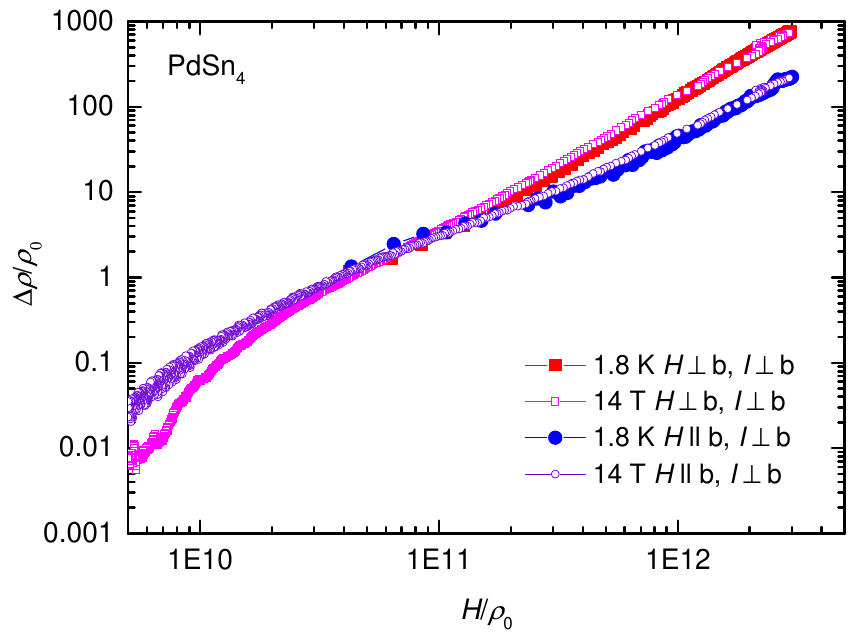}%
	\caption{(color online) Kohler's plot of PdSn$_4$ magnetoresistance data. Red filled squares are based on $\rho(H)$ data taken at 1.8 K.($H\,\perp\,b$, $I\,\perp\,b$) Magenta open squares are based on $\rho(T)$ data taken under the magnetic field of 140 kOe. ($H\,\perp\,b$, $I\,\perp\,b$) Blue filed circles are based on $\rho(H)$ data taken at 1.8 K.($H\,\perp\,b$, $I\,\parallel\,b$) Purple open circles are based on $\rho(T)$ data taken under the magnetic field of 140 kOe.($H\,\perp\,b$, $I\,\parallel\,b$) 
		\label{Kohler}}
\end{figure}

\begin{figure}
	\includegraphics[scale=1]{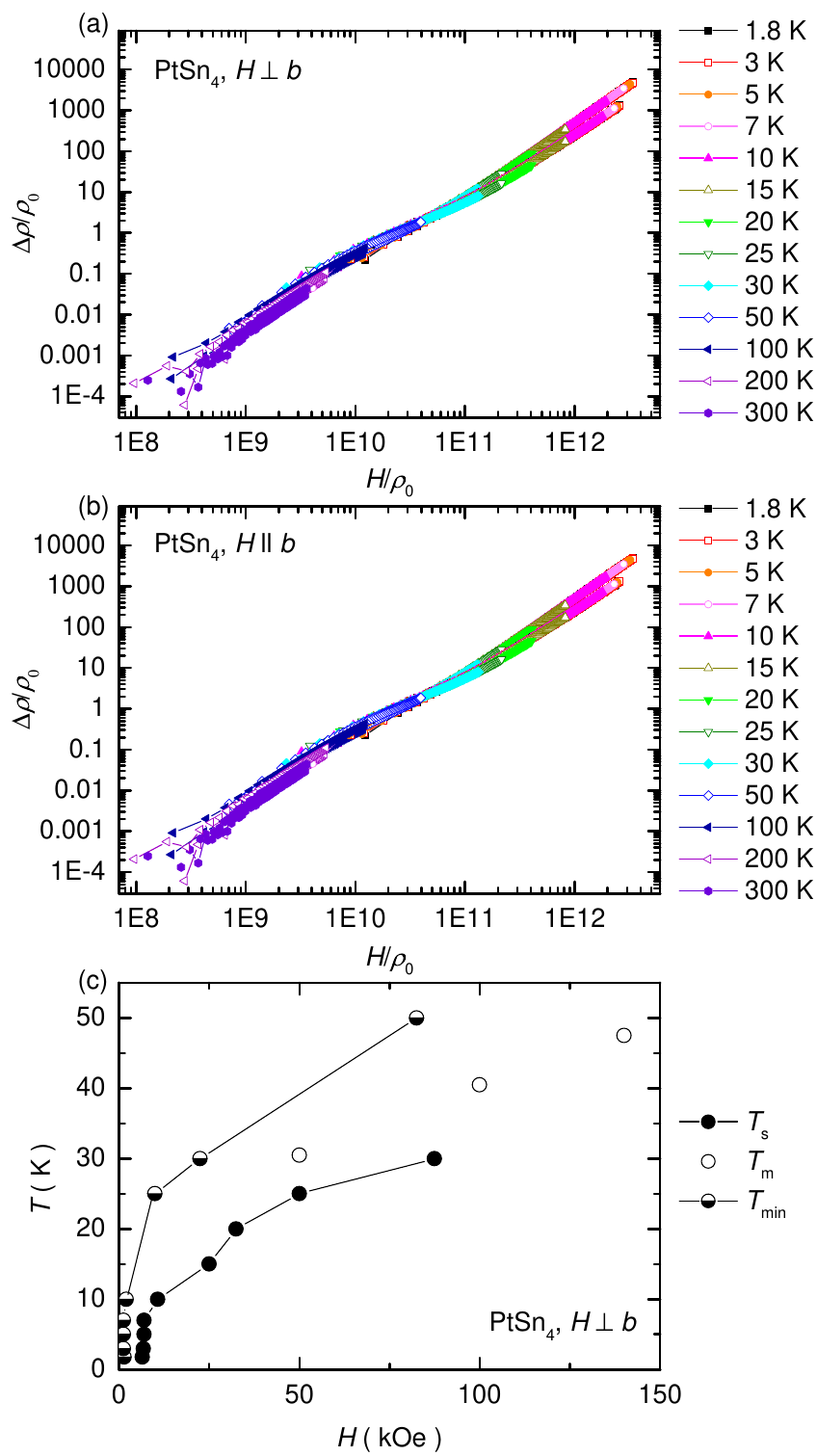}%
	\caption{(color online) Kohler's plot of PtSn$_4$ magnetoresistance data with various temperatures; 1.8\,K, 3\,K, 5\,K, 7\,K, 10\,K, 15\,K, 20\,K, 25\,K, 30\,K, 50\,K, 100\,K, 200\,K, 300\,K with two different magnetic field directions from ref.\,\onlinecite{Mun2012} (a) $H\,\perp\,b$ and (b) $H\,\parallel\,b$. (c) Temperature-field phase diagram for PtSn$_4$ for $H\,\perp\,b$. Filled black circles are the starting point of the Kohler's rule slope change in (a) in the high magnetic field regime. Open black circles are the temperature of the local minimum in $\rho(T)$ data for an applied magnetic field of 50, 100 or 140\,kOe. Half filled circles are the shallowest slope in the Kohler's plot in (a).  
		\label{phase}}
\end{figure}

Let us now investigate whether a classical description of the electronic motion can give insight into the MR results. The well-known Kohler's rule states that the change of resistivity $\Delta\rho$, where $\Delta\rho\,=\,\rho(T,H)-\rho_{0}$ and where $\rho_{0}\,\equiv\,\rho(T,H\,=\,0)$, under the magnetic field $H$ can be described as a scaling function of the variable $H\tau$.\cite{ziman1960} Here, $\tau$ is the mean time between scattering events of conduction electrons, and it is inversely proportional to $\rho_0$. Such a scaling behavior expresses a self-similarity of the electronic orbital motion across different length scales: an invariance of the magnetoresponse under the combined transformation of shrinking the orbital length $L \sim 1/H$ by increasing $H$ while at the same time also increase the scattering rate $1/\tau$ by the same factor such that $H \tau$ remains unchanged, indicates that the system behaves identical on different length scales. The formula for the Kohler's rule can be written as follows with scaling function $F(x)$. 
\begin{equation}
\Delta\rho/\rho_{0}\,=\,F(H/\rho_{0})
\end{equation} 
It should be appreciated that Kohler's rule is both powerful and simplified; Kohler's rule scaling breaks down if different scattering mechanisms emerge on different temperature or lengthscales or if Landau orbit quantization plays a role.\cite{Pippard1989}

In Figs.\,\ref{Kohler} and \ref{phase} we plot MR data for PdSn$_4$ and PtSn$_4$ respectively, using the scaling ansatz of Eq.\,(5) that emphasizes the Kohler’s rule behavior found for both of these compounds. In Fig.\,\ref{Kohler}, for PdSn$_4$ we plot MR data, for two field directions, extracted from two different extremes of MR sweeps: $\rho(T\,=\,1.8$\,K, 500\,Oe\,$\leq\,H\,\leq140$\,kOe) and $\rho$(1.8\,K\,$\leq\,T\,\leq\,300$\,K, $H\,=\,140$\,kOe).  Remarkably although the two data sets, lie in very different field and temperature regimes, they fall precisely on top of each other, describing a sigmoidal-shaped curve. These plots demonstrate that the behavior of all of the MR data for PdSn$_4$, from 1.8\,K to 300\,K, from 0\,Oe to 140\,kOe is captured and described by a Kohler's rule. Figure\,\ref{phase} shows the same generalized Kohler's plot for MR data collected on PtSn$_4$.\cite{Mun2012} We find that all of the $\rho(T, 500\,\textrm{Oe}\,\leq\,H\,\leq\,140$\,kOe) data for temperatures ranging from 1.8\,K all the way up to 300\,K fall onto similar sigmoidal-shaped curves.  

The importance of the such complete MR data collapse in Figs.\,\ref{Kohler} and \ref{phase} cannot be over emphasized. In Fig.\,\ref{Kohler}, for PdSn$_4$, we find precisely the same behavior at 1.8\,K as we so for temperatures ranging up to 300\,K. The two very different cuts through $T$-$H$ phase space yield the same Kohler's rule curve. This means that the behavior of the MR at 1.8\,K is governed by the simple physics as the behavior at 300\,K; i.e. there is not a “metal-to-insulator” transition or other such feature associated with the local minima in the $\rho(T)$ data for any given applied field.  Instead, given that a generalized form of Kohler's rule describes the PdSn$_4$ and PtSn$_4$ data exceptionally well, any features that are seen in the $\rho(T,H)$ data in the low temperature/high field limit originate from the same physics giving rise to the high temperature/low field behavior.

This generalized Kohler's rule analysis, of course, raises the question of what is the origin of the sigmoidal-shape we find for both PdSn$_4$ and PtSn$_4$? In both Figs.\,\ref{Kohler} and \ref{phase} there is a well defined, linear, high field/low temperature (upper right hand corners of the plots) region that can be associated with power law behavior. As either temperature is increased or field is lowered the curvature associated with the sigmoidal-shape appears and grows. It is possible, even likely, given the clear dHvA oscillations, that the high magnetic field (or low temperature) regime is where $\omega_{c}\tau\,>\,1$. Conversely, high temperatures and low magnetic field will be where $\omega_{c}\tau\,<\,1$. It should be noted that large MR can be expected only when $\omega_{c}\tau\,\gg\,1$.\cite{Pippard1989} For both PtSn$_4$ and PdSn$_4$ the middle of the sigmoidal region of the curves occur for $\Delta\rho/\rho_{0}\,\sim\,1$; this may provide a physical sense of, or criterion for, $\omega_{c}\tau\,\gg\,1$.

We can use the Kohler's rule data shown in Fig.\,\ref{phase} (a) to delineate the low temperature/high field part of the $T$-$H$ phase space from the high temperature/low field part. In Fig.\,\ref{phase}  (c) we plot data inferred from (i) the deviation from high field/low temperature linear behavior in the upper right part of Fig.\,\ref{phase} (a), and (ii) the position of the inflection point associated with the sigmoidal-shaped curve. These data provide an estimate of the extent of the high field/low temperature region for this specific PtSn$_4$ sample. We can also use the local minima in the $\rho(T)$ data measured for a constant applied magnetic field to define a set of ($T$,$H$) data points. We include these data in Fig.\,\ref{phase} (c) as well; they clearly are associated with crossing over from a low temperature limit to a high temperature limit in terms of $\omega_{c}\tau$.

The challenge that PdSn$_4$ and PtSn$_4$ present is to understand what specific feature of their band structure is responsible for the remarkable size of the MR that is observed. This still unresolved point is what makes research into these compounds, as well as other XMR materials a compelling topic for ongoing research.

\section{Conclusion}

In conclusion, the thermodynamic and transport properties of PdSn$_4$ are strikingly similar to those of PtSn$_4$; both compounds have closely related temperature dependent specific heat, anisotropic resistivity and magnetization, although the precise magnitudes of RRR and MR are quantitatively different. We have focused on the origin of XMR in these two materials. Based on our analysis, we can rule out carrier compensation and the surface state Dirac arc node features observed in PtSn$_4$ can be considered as key reasons for the XMR. Instead, we find that, by using a generalized Kohler's plot, all of the MR data for both PdSn$_4$ and PtSn$_4$ collapse onto simple sigmoidal manifolds. For PdSn$_4$, the fact that MR data from an isothermal field sweep at 1.8\,K and a constant field (140\,kOe) temperature sweep from 1.8\,K to 300\,K fall indistinguishably on top of each other (with similar results for PtSn$_4$) clearly emphasizes that the behavior of the MR data for these materials is captured by the basic metal physics encoded into Kohler's rule. A future challenge is to identify the specific property of PdSn$_4$, PtSn$_4$ and other XMR compounds that give rise to the remarkable size and durability of the observed XMR.

Note that as we were preparing to submit this work, another paper on PdSn$_4$ was posted.\cite{Xu2017} Single crystals of PdSn$_4$ were grown out of excess Sn, but instead of separating the crystals from excess Sn by decanting in the liquid state, the excess Sn was allowed to solidify around the PdSn$_4$ crystals. PdSn$_4$ crystals were ultimately separated from solidified Sn flux by acid etching. The samples reported in ref.\,\onlinecite{Xu2017} have substantially lower RRR values as well as substantially reduced MR.  Although ref.\,\onlinecite{Xu2017} only reports a subset of the quantum oscillation frequencies we were able to detect, the ones that are reported agree with what we found to a large degree. Given the importance of $\rho_{0}$ in the physics described by Kohler's rule, it is probably preferable to compare PdSn$_4$ and PtSn$_4$ samples grown in similar ways and with closer $\rho_{0}$ and RRR values.

\begin{acknowledgements}
We thank E. Kunz Wille for helpful dicussion. Research was supported by the U.S. Department of Energy, Office of Basic Energy Sciences, Division of Materials Sciences and Engineering. Ames Laboratory is operated for the U.S. Department of Energy by the Iowa State University under Contract No. DE-AC02-07CH11358. Na Hyun Jo and Soham Manni are supported by the Gordon and Betty Moore Foundation EPiQS Initiative (Grant No. GBMF4411). Yun Wu and Lin-Lin Wang are supported by Ames Laboratory's Laboratory-Directed Research and Development (LDRD) funding. Peter P. Orth acknowledges support from Iowa State University Startup Funds. 
\end{acknowledgements}

\clearpage

\section*{References}
%\bibliographystyle{apsrev4-1}
%\bibliography{references}

\begin{thebibliography}{59}
\bibitem [{com()}]{commentary}%
\BibitemOpen
\href@noop {} {\bibinfo  {journal} {For example, similarly good fits to the
		data could be found for $n_{e}\,\simeq\,n_{h}\,\sim\,10^{33}$ m$^{-3}$ with $c_{\mu}$\,=\,$\mu_{h}/\mu_{e}$\,=\,0.0013. Clearly the size of $n_{e}$ and $n_{h}$ is not physical, but it is worth nothing that they are still roughly equal}\ }\BibitemShut {NoStop}%
\end{thebibliography}

%

\clearpage

\end{document}